\RequirePackage{silence}
\documentclass[10pt,aps,prd,twocolumn,superscriptaddress]{revtex4-2}

\usepackage{amsmath}
\usepackage{amssymb}
\usepackage{graphicx}
\usepackage{dcolumn}
\usepackage{bm}
\usepackage{anyfontsize}
\usepackage{xcolor}
\usepackage[normalem]{ulem} 
\usepackage{booktabs,multirow,tabularx}
\usepackage{aas_macros}

\usepackage{hyperref}

\begin{document}

\preprint{APS/123-QED}

\title{Revealing galaxy assembly bias with luminosity-weighted marked correlations}

\author{Joaquin Armijo}
\email{joaquin.armijo@usp.br}
\affiliation{Department of Mathematical Physics, Institute of Physics, University of São Paulo, R. do Matão 1371, 05508-090, São Paulo, SP, Brazil}
\affiliation{Departamento de Física, Universidad Técnica Federico Santa María, Avenida Vicuña Mackenna 3939, San Joaquín, Santiago, Chile}

\author{Roberta Vasallo}
\affiliation{Department of Mathematical Physics, Institute of Physics, University of São Paulo, R. do Matão 1371, 05508-090, São Paulo, SP, Brazil}

\author{L. Raul Abramo}
\affiliation{Department of Mathematical Physics, Institute of Physics, University of São Paulo, R. do Matão 1371, 05508-090, São Paulo, SP, Brazil}
\author{Antonio D. Montero-Dorta}
\affiliation{Departamento de Física, Universidad Técnica Federico Santa María, Avenida Vicuña Mackenna 3939, San Joaquín, Santiago, Chile}

\begin{abstract}
We investigate whether luminosity-weighted marked correlation functions can reveal secondary halo information beyond that encoded by halo mass alone, as expected in scenarios of galaxy assembly bias. We propose a controlled null test in which galaxy luminosities follow the same mean luminosity--halo mass relation and scatter, while their residuals are either (A) uncorrelated with secondary halo properties or (B) correlated with a standardized secondary halo property, $s$. Both models have identical halo occupation, galaxy positions, luminosity function, and unweighted projected two-point correlation function, $w_p(r_p)$, so their distinction is encoded entirely in the luminosity marks. At leading order, the projected marked correlation function difference, $\Delta\mathcal{M}_p(r_p)$, scales as $\rho^2\sigma_{\log L}^2\xi_{ss}(r_p)$, where $\rho$ controls the coupling between the luminosity and secondary-property residuals, $\sigma_{\log L}$ is the scatter in log luminosity at fixed halo mass, and $\xi_{ss}$ describes the spatial correlation of the secondary property. We test this framework using a DESI-BGS mock catalogue over $0.1<z<0.4$. For the injected dependence, we detect a marked-clustering deviation over $0.11<r_p/(h^{-1}\,{\rm Mpc})<1.04$, obtaining $\chi^2=30.14$ relative to the null covariance. A one-parameter test rejects the mass-only hypothesis at greater than $5\sigma$ significance ($p=2.13\times10^{-7}$) for $|\rho|>0.63$. At fixed $\sigma_{\log L}=0.23$, correlations as weak as $|\rho|\simeq0.52$ remain detectable, with the results stable under variations in the HOD prescription and modelled galaxy selection effects. These results establish luminosity-weighted marked clustering as a sensitive null test for property-based galaxy assembly bias and secondary halo information hidden from unweighted two-point clustering.
\end{abstract}

\maketitle


\section{Introduction} \label{sec:introduction}

\noindent In the standard cosmological model, $\Lambda$CDM, the large-scale structure of the Universe originates from small primordial matter-density fluctuations that grow through gravitational instability \citep{Bernardeau2002,AnguloHahn2022}. As structure evolves hierarchically, matter collapses into dark matter haloes, which provide the underlying framework within which galaxies form and evolve \citep{PressSchechter74,WhiteRees78}. The properties and spatial distribution of galaxies are expected to be closely connected to those of their host haloes, a relation commonly referred to as the galaxy--halo connection \citep[for a comprehensive review, see][]{WechslerTinker2018}. Empirical descriptions of this connection, including halo occupation distribution (HOD) and conditional luminosity function (CLF) models, have been remarkably successful under the standard assumption that halo mass is the primary variable controlling the galaxy population \citep{Berlind2002,Yang2003}. In this picture, galaxy abundance and spatial clustering are determined by how galaxies populate haloes of different masses, while on large scales galaxy clustering can be related to the underlying matter distribution through galaxy bias \citep{DesjacquesJeongSchmidt2018}. Because halo bias depends strongly on halo mass, the distribution of host halo masses provides a natural connection between galaxy abundance and galaxy clustering \citep{MoWhite1996,ShethMoTormen2001,Tinker2010}.

A central open question is whether halo mass provides a complete description of the galaxy--halo connection \citep{Zentner2014,WechslerTinker2018}, and what summary statistics can be used to reveal any secondary dependencies. Dark matter haloes of the same mass can exhibit different formation times, concentrations, angular momenta, accretion histories, and environmental information \citep{Wechsler2006,Bett2007,Fakhouri2010}. Several of these secondary properties are themselves correlated with halo clustering at fixed mass, an effect commonly referred to as halo assembly bias or secondary halo bias \citep{Gao2005,Wechsler2006}. At fixed halo mass, if either the properties of galaxies or the number of galaxies hosted by a halo depend on secondary halo properties, this dependence can propagate to the galaxy population and is referred to as galaxy assembly bias \citep{Hearin2016,WechslerTinker2018}. The dependence of halo occupation on secondary halo properties at fixed mass is commonly called occupancy variation; coupled with halo assembly bias, it can generate an occupation-based galaxy assembly-bias signal \citep{Zehavi2018,Artale2018,Contreras2019}. Another important question is whether residuals in galaxy properties at fixed halo mass retain information about secondary halo properties beyond a main relation, and whether these correlations leave an observable imprint on galaxy clustering \citep{CrotonGaoWhite2007,HearinWatsonDouglas2013,WechslerTinker2018}. Hydrodynamical simulations support this property-based behaviour, showing that halo formation and mass-accretion histories can be reflected in observable galaxy properties and in secondary differences in galaxy clustering at fixed halo mass \citep{MonteroDorta2020,MonteroDorta2021}. Such information may be encoded not only in the occupation of haloes by galaxies, but also in observable galaxy properties---including luminosity, colour, and star-formation activity---at fixed halo mass \citep{HearinWatsonDouglas2013,WechslerTinker2018}.

Observational constraints on galaxy assembly bias remain mixed and depend on the galaxy sample, the secondary halo dependence being tested, and the observables and modelling assumptions employed. Combining counts-in-cylinders with projected clustering, analyses of luminosity-threshold samples from SDSS DR7 and the DESI BGS One-Percent Survey have found evidence for central occupation assembly bias, particularly for some lower-luminosity samples \citep{Wang2022,Pearl2024}. A model-based analysis of DESI emission-line galaxy clustering has also inferred a non-zero assembly-bias signal and examined its connection to galaxy conformity \citep{Ortega-Martinez2025}. Conversely, an analysis of SDSS galaxies found no statistically significant central or satellite occupation assembly bias in its fiducial constraints \citep{Salcedo2022}, while a direct measurement using DESI DR1 BGS galaxies found satellite occupation assembly bias consistent with zero \citep{Shao2025}. These differing results motivate complementary probes of secondary halo dependence, including tests based on observable galaxy properties rather than on galaxy occupation alone.

Marked correlation functions provide a natural way to test for secondary dependencies in the galaxy--halo connection. Rather than weighting all galaxy pairs equally, a marked statistic assigns each galaxy a mark associated with a chosen property and weights each pair by the product of its marks, thereby measuring how that property correlates with pair separation and environment
\citep{Beisbart2000,ShethTormen2004,Sheth2005}. For example, by using halo formation time, \citet{ShethTormen2004} applied marked correlation functions to probe the environmental dependence of halo formation, providing an early application of marked statistics to secondary halo information. Suitably chosen marks can distinguish models that are degenerate in conventional unweighted two-point clustering; for example, \citet{WhitePadmanabhan2009} used a local-density mark to break a degeneracy between the HOD and cosmology. More generally, nonlinear density marks can make two-point-like statistics sensitive to information contained in higher-order correlations \citep{PhilcoxMassaraSpergel2020}. For luminosity marks, \citet{Skibba2006} showed that a non-zero marked signal is expected even in a standard mass-only galaxy--halo model. This signal arises because the mean luminosity of galaxies depends on halo mass, while massive haloes preferentially populate denser environments \citep{ShethTormen2004,Sheth2005,Skibba2006}. Consequently, luminosity and environment can be correlated even in the absence of any secondary halo dependence at fixed mass. Luminosity-weighted marked correlations can provide information complementary to the ordinary two-point correlation function and offer a continuous measure of the relation between galaxy properties and environment without requiring the galaxy population to be divided into discrete environmental classes
\citep{ShethTormen2004}.

In this work, we build on the mass-only interpretation of luminosity-weighted marked clustering and ask whether it can reveal secondary halo information beyond that encoded by halo mass alone. Throughout this work, we use the term galaxy assembly bias specifically to describe a dependence of observable galaxy properties on secondary halo properties at fixed halo mass \citep{WechslerTinker2018}. We construct a controlled null test in which the mean luminosity--halo mass relation and its scatter define the mass-only prediction, and test whether correlating luminosity scatter with a spatially correlated secondary halo property generates an additional marked-clustering contribution. The halo occupation and galaxy positions are held fixed, so the unweighted two-point clustering remains unchanged, while galaxy luminosities can retain information about the secondary halo property. Then, our test targets a property-based manifestation of galaxy assembly bias rather than an occupation-based one. The resulting difference in the luminosity-weighted marked correlation function provides a controlled test of secondary halo information that is invisible to standard unweighted two-point clustering.

We organize this paper as it follows: In Section \S \ref{sec:LumMarked} we explain the methodology of the Luminosity-weighted marked correlation function. We define the property-based galaxy assembly bias and design the null test in Section \S \ref{sec:NullTest} . The simulations and the elaboration of the mock catalogue to perform the test are explained in Section \S \ref{sec:Sims}. We show the result of the  application of the model and test for the mocks in Section \S \ref{sec:Results}. We add the conclusions of this work ins Section \S \ref{sec:conclusions}.

\section{Luminosity-weighted marked correlations} \label{sec:LumMarked}

\begin{figure*}
\centering
\includegraphics[width=0.65\linewidth]{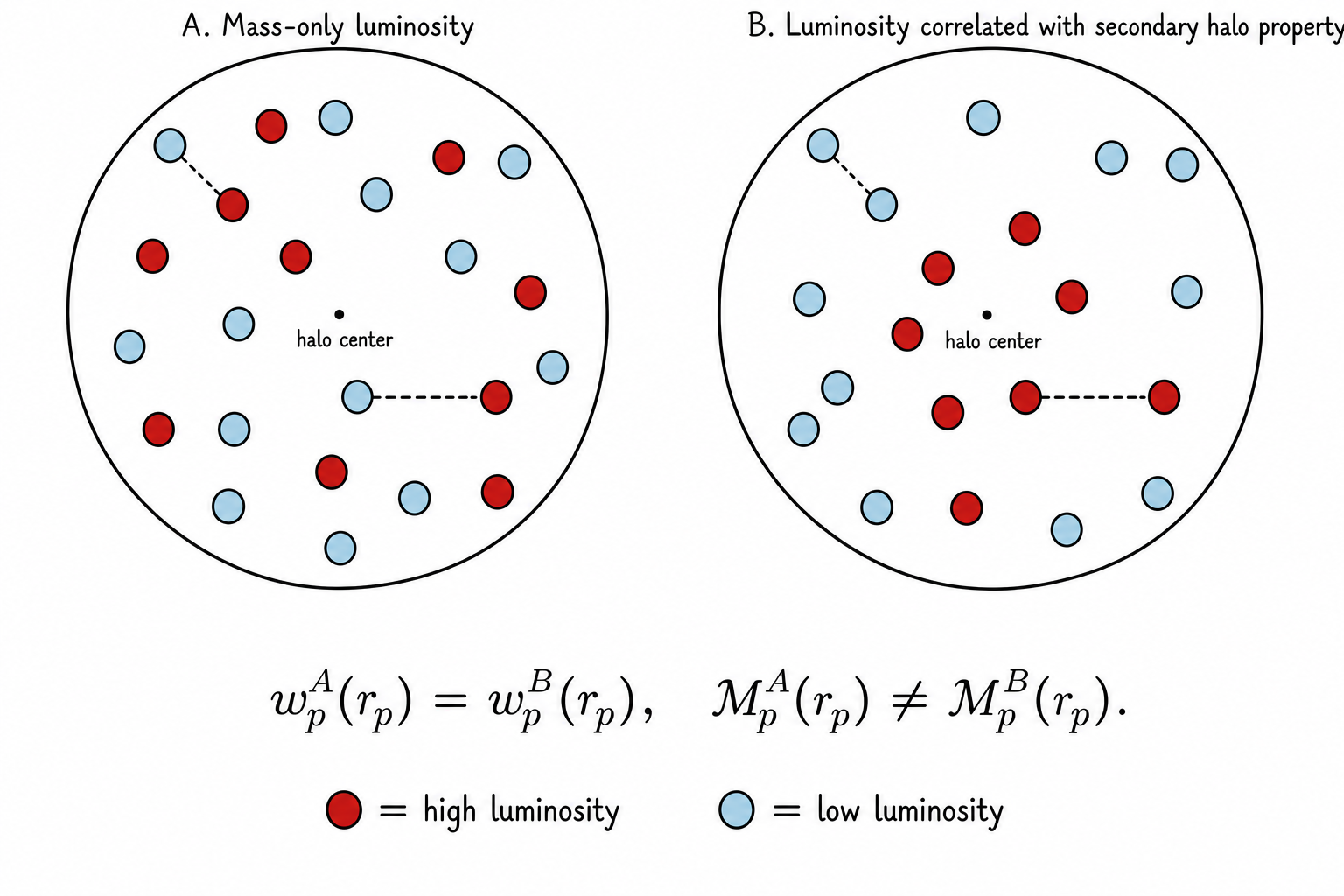}
\caption{Schematic illustration of the two luminosity-assignment models used in the null test. Left: In Model A, galaxy luminosities follow a halo-mass-only relation, and luminosity residuals (galaxy luminosity $-$ mean luminosity) are assigned independently once halo mass is specified, colouring high (red) and low (light-blue) luminosity galaxies. Right: In Model B, the same luminosities are reassigned so that their residuals correlate with a secondary halo property. For visual clarity, this correlation is represented by placing the most luminous galaxies preferentially closer to the halo centre; the radial ordering is illustrative and can represent any property, such as concentration, formation history, or environment. The halo occupation and galaxy positions are identical in the two models, so their unweighted projected two-point correlation functions are the same. Their luminosity-weighted marked correlation functions can nevertheless differ because the luminosity marks contain different spatially correlated information.}
\label{fig:halo_luminosity_weighted_dependency}
\end{figure*}

The luminosity-weighted marked correlation functions provide a way of measuring how galaxy luminosity is distributed over the galaxy density field without splitting the sample into separate luminosity bins. Following the halo-model interpretation of \cite{Cooray2002} and \cite{Skibba2006}, the marked statistic can be written as the ratio between a luminosity-weighted correlation function and the ordinary two-point correlation function, schematically:

\begin{equation}
    \mathcal{M}(r) = \frac{1+W(r)}{1+\xi(r)}.
\end{equation}

Here, $W(r)$ is obtained from the same galaxy pair counts that define $\xi(r)$, but with each galaxy weighted by its luminosity mark. Considering the halo model, both $\xi$ and $W$ can be decomposed into one-halo and two-halo terms; the difference is that the weighted statistic replaces the ordinary halo occupation moments by luminosity-weighted occupation moments. Therefore, even in a model with no explicit environmental dependence at fixed halo mass, a luminosity-marked signal is expected whenever the mean luminosity of galaxies depends on halo mass, because massive haloes preferentially populate dense environments \citep{ShethTormen2004,Sheth2005}. This provides a standard mass-only prediction for luminosity-marked clustering \citep{Skibba2006}, against which we test additional secondary-property dependence by introducing a coupling between the luminosity scattering and a secondary halo property.

In the standard halo occupation picture, galaxy luminosity is primarily determined by the mass of its host halo. We describe the luminosity distribution at fixed halo mass using a conditional luminosity function (CLF), $\Phi(L|M)$ \citep{Yang2003}. The mean number of galaxies hosted by a halo of mass $M$ above a luminosity threshold is
\begin{equation}
\left\langle N(M) \right\rangle =
\int_{L_{\rm min}}^{\infty}
\Phi(L|M)\,{\rm d}L .
\label{eq:CLF}
\end{equation}

The CLF specifies the distribution of galaxy luminosities at fixed
host-halo mass determining the mean occupation above a
luminosity threshold. To describe the luminosity of an individual
galaxy, we write
\begin{equation}
\log L_i =
\log \bar{L}(M_i)
+\sigma_{\log L}\epsilon_i,
\label{eq:luminosity_model}
\end{equation}
where $\log\bar{L}(M_i)\equiv\langle\log L_i\mid M_i\rangle$ is the
mean log-luminosity--halo mass relation, $\sigma_{\log L}$ is the
standard deviation of $\log L$ at fixed halo mass, expressed as a natural logarithm, and
$\epsilon_i\sim\mathcal{N}(0,1)$ is a standard-normal distribution. 

We define the normalized luminosity mark
\begin{equation}
m_i = \frac{L_i}{\langle L\rangle},
\label{eq:luminosity_mark}
\end{equation}
where $\langle L\rangle$ is the mean luminosity of the galaxy sample
under consideration. By construction, $\langle m\rangle=1$. In
general, $\langle L\rangle$ differs from $\bar{L}(M)$: it is the
sample-averaged luminosity across the host-halo mass distribution.

At fixed halo mass, we can decompose the mark into its conditional mean and
residual,
\begin{equation}
m_i = \langle m_i\mid M_i\rangle + \delta m_i,
\label{eq:mark_decomposition_single}
\end{equation}
where
\begin{equation}
\langle \delta m_i\mid M_i\rangle=0.
\label{eq:residual_zero_mean}
\end{equation}

The marked correlation function is the mean product of the marks for galaxy pairs separated by $r$,
\begin{equation}
\mathcal{M}(r)
\equiv
\left\langle m_i m_j \right\rangle_r .
\label{eq:marked_correlation_definition}
\end{equation}

Using the decomposition above, the marked correlation function can be written as
\begin{align}
\mathcal{M}(r)
={}&
\left\langle
\left\langle m|M_i\right\rangle
\left\langle m|M_j\right\rangle
\right\rangle_r
\nonumber\\
&+
\left\langle
\left\langle m|M_i\right\rangle
\delta m_j
\right\rangle_r
+
\left\langle
\left\langle m|M_j\right\rangle
\delta m_i
\right\rangle_r
\nonumber\\
&+
\left\langle
\delta m_i \delta m_j
\right\rangle_r .
\label{eq:marked_correlation_decomposition}
\end{align}

We define the null hypothesis of this work as the case in which, once halo mass is specified, galaxy luminosity is described by the mean luminosity--mass relation plus stochastic scatter. Under this hypothesis, the scatter contains no additional spatially correlated information: luminosity residuals are independent of pair separation and the properties of the neighbour galaxies inside a halo. The conditional-mean decomposition in Eq.~\eqref{eq:marked_correlation_decomposition} then has vanishing cross terms, while the residual-correlation term $\left\langle\delta m_i\delta m_j\right\rangle_r$ also vanishes for distinct galaxy pairs. The marked correlation is therefore determined by the mass-dependent mark completely. To test for information beyond halo mass without changing the mean luminosity--mass relation or its scatter, we construct an alternative model in which the luminosity residuals are correlated with a secondary halo property at fixed mass. If this property is spatially correlated (see Figure~\ref{fig:halo_luminosity_weighted_dependency}), the luminosity residuals inherit that correlation and contribute to the marked-clustering signal. As illustrated in Figure~\ref{fig:halo_luminosity_weighted_dependency}, the two models have almost identical halo occupation and galaxy positions, and subsequently the same unweighted two-point clustering, but differ in the spatial assignment of their luminosity marks.

\subsection{Mass-only luminosity model}

We define a baseline model, denoted Model A, in which the luminosity residual is independent of any secondary halo property,
\begin{equation}
\epsilon_i^{\rm A} = \eta_i,
\qquad
\eta_i \sim \mathcal{N}(0,1).
\label{eq:model_A}
\end{equation}

For two distinct galaxies,
\begin{equation}
\left\langle \eta_i \eta_j \right\rangle = 0,
\qquad i \neq j ,
\label{eq:eta_uncorrelated}
\end{equation}
the luminosity scatter produces stochastic variation in the marks characterised only by $\sigma_{\log L}$, but introduces no additional spatially correlated luminosity information.

The systematic dependence of the luminosity-weighted clustering is then determined entirely by halo mass. Model A provides the mass-only reference against which the alternative model is compared.

\subsection{Dependence on a secondary halo property}

\begin{figure}
\centering
\includegraphics[width=0.85\linewidth]{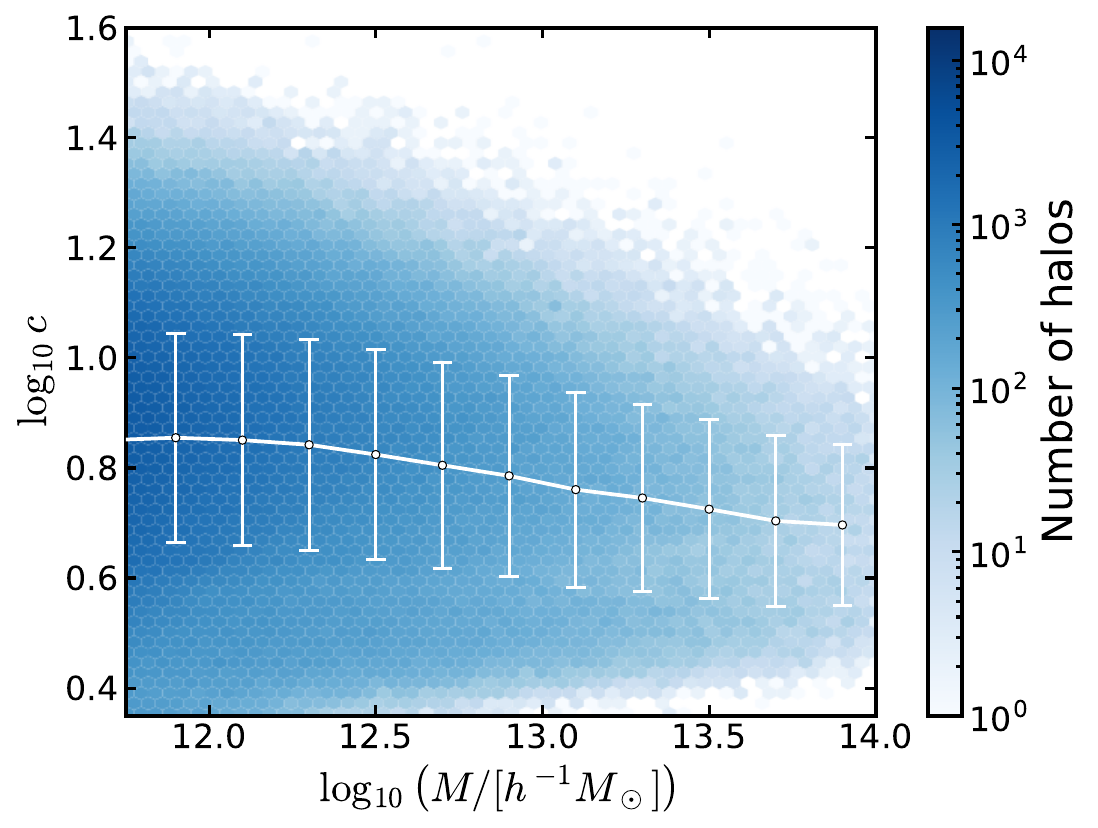} \\
\includegraphics[width=0.85\linewidth]{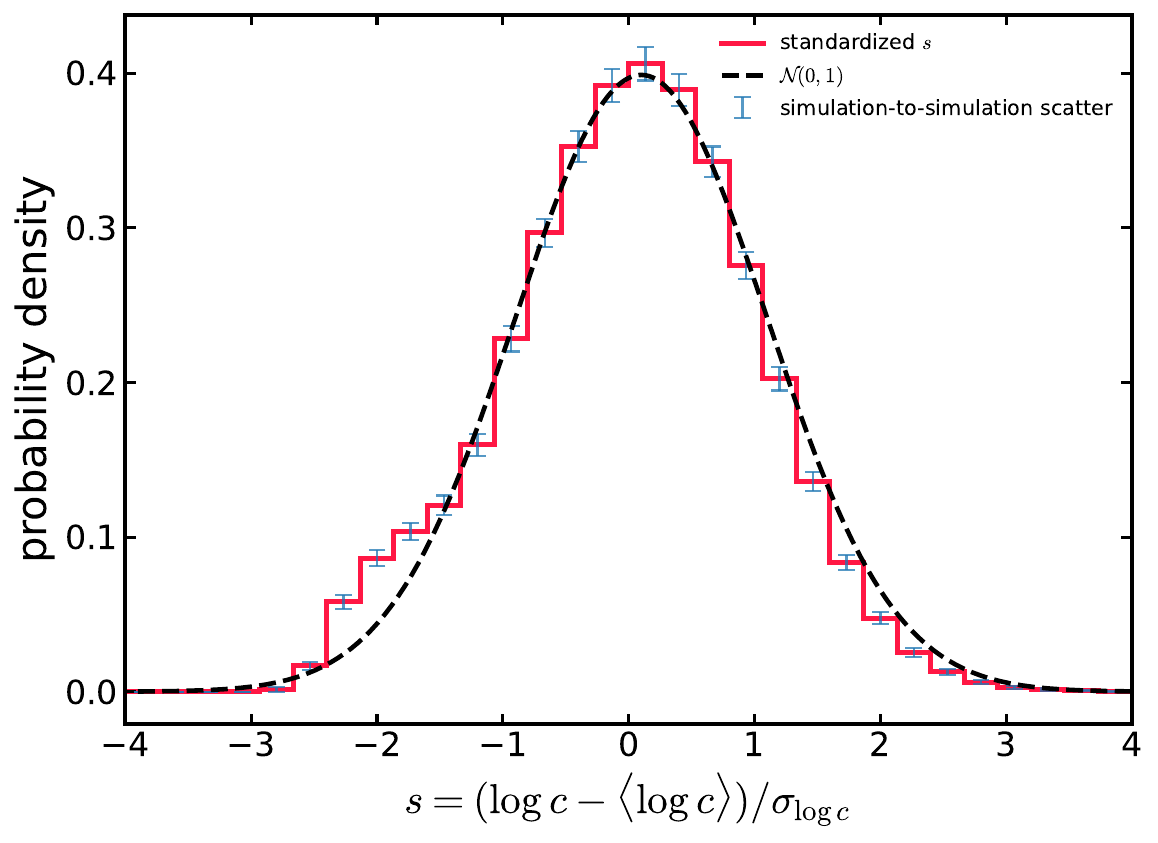}
\caption{Example of the standardised mass dependent property displayed by the concentration--mass relation. Top: We bin log-concentration, $\log c$, from the \texttt{abacus} simulation inferred by using the $r_{v_{\max}}$ as a proxy, for several bins in the log halo mass $M_{200m}$ (blue hexagons), with a gradient colour indicating the Number of haloes (colour bar). In this example, we show 11 bins, between $  11.8 <\log M/(h^{-1}\ M_{\odot}) < 14.0$, which represents the halos where the relation can be standardised by calculating the mean $\langle\log c\rangle$ and variance $\sigma_{\log c}^2$ (white line with error bars). We display only the mass bins of haloes with $N_{\rm p} \gtrsim 1000$, where concentration is numerically well defined for a halo. Bottom: Probability density function $s$ (red histogram) defined by $\log c$ for the mass bins showed in the top panel. This shows that $s$ for a fix halo mass, can be expressed as a unit Gaussian distribution (dashed line). We add Poisson error (light-blue error bars) bars to show the scatter of the stacked 500 simulations we arbitrarily select from the AbacusSummit small suite.}
\label{fig:concentration_standardization}
\end{figure}

We next consider the possibility that galaxy luminosity contains information about a secondary halo property at fixed halo mass \citep{Lehmann2017,Oyarzun2024}, as sketched in Figure~\ref{fig:halo_luminosity_weighted_dependency}, showing a luminosity dependent clustering. For a generic halo property $q$, we define a standardized residual
\begin{equation}
s_i =
\frac{
q_i-\left\langle q|M_i\right\rangle
}{
\sigma_q
}.
\label{eq:secondary_property_standardization}
\end{equation}

By construction,
\begin{equation}
\left\langle s|M \right\rangle \simeq 0,
\qquad
{\rm Var}(s|M) \simeq 1.
\label{eq:s_standardization_conditions}
\end{equation}

For halo concentration, for example, we take $q=\log c$, such that
\begin{equation}
s_i =
\frac{
\log c_i-\left\langle\log c\mid M_i\right\rangle
}{
\sigma_{\log c}(M_i)
}.
\label{eq:concentration_standardization}
\end{equation}

As shown in Figure~\ref{fig:concentration_standardization}, this procedure removes both the mean concentration--mass relation and the mass dependence of its scatter. The secondary dependence introduced below does not arise from the ordinary correlation between halo mass and concentration. Coupling luminosity directly to concentration could modify the mean luminosity--mass relation and, consequently, the resulting luminosity function \citep{Lehmann2017}. Instead, our construction correlates luminosity residuals with concentration residuals at fixed halo mass, thereby preserving the mean luminosity--mass relation. We adopt concentration as an illustrative secondary property because it can be robustly standardized at fixed mass \citep{LopezCano2022}, and exhibits spatially correlated variations associated with secondary halo bias. Nevertheless, this definition is not specific to concentration, but to any standard secondary halo property.

In Model B, the luminosity residual is correlated with the secondary halo property according to
\begin{equation}
\epsilon_i^{\rm B}
=
\rho s_i
+
\sqrt{1-\rho^2}\,\eta_i,
\label{eq:model_B}
\end{equation}
where $\rho$ is the correlation coefficient, which controls the strength of the luminosity--secondary-property residuals coupling. The same realization of $\eta_i$ is used in Models A and B, so that their difference is driven only by the additional dependence on $s$.

If $s$ is standardized, the variance of $\epsilon^{\rm B}$ is approximately preserved,
\begin{equation}
\left\langle
\left(\epsilon^{\rm B}\right)^2
\right\rangle
\simeq
\rho^2 + 1-\rho^2 = 1.
\label{eq:model_B_variance}
\end{equation}

Models A and B can be constructed with the same halo catalogue, galaxy occupation, galaxy positions, mean luminosity--mass relation, and luminosity-scatter amplitude. Their intended difference is only the correlation between luminosity scatter and the standardised residuals.

For two distinct galaxies, the independent random contributions from $\eta$ vanish in the ensemble average, while the component proportional to the secondary property remains correlated. Therefore, we obtain
\begin{equation}
\left\langle
\epsilon_i^{\rm B}
\epsilon_j^{\rm B}
\right\rangle_r
=
\rho^2
\left\langle s_i s_j\right\rangle_r .
\label{eq:epsilon_B_correlation}
\end{equation}

We define
\begin{equation}
\xi_{ss}(r)
\equiv
\left\langle s_i s_j\right\rangle_r ,
\label{eq:xi_ss}
\end{equation}
which corresponds to the marked correlation function of the standardized secondary halo property at separation $r$. This quantity describes the scale-dependent spatial correlation of the secondary-property residuals after their mean dependence on halo mass has been removed.

The importance of Eq.~\eqref{eq:epsilon_B_correlation} becomes clear when considering the decomposition of the luminosity mark in Eq.~\eqref{eq:marked_correlation_decomposition}. The mass-dependent contribution, $\langle m|M\rangle$, is the same by construction in Models A and B, since both models share the same mean luminosity--mass relation. The additional dependence introduced in Model B is entirely contained in the residual component $\delta m$. Consequently, when comparing the marked correlations of the two models, the term of interest is the residual correlation $\langle\delta m_i\delta m_j\rangle_r$, whose scale dependence is inherited from $\xi_{ss}(r)$.

Since the luminosity scatter in Eq.~\eqref{eq:luminosity_model} is defined in natural-log space, the luminosity mark depends exponentially on the residual,
\begin{equation}
L_i =
\bar{L}(M_i)
\exp\left(\sigma_{\log L}\epsilon_i\right).
\label{eq:luminosity_exponential_form}
\end{equation}

Expanding this dependence to leading order, as done in Eq.~\eqref{eq:mark_decomposition_single}, shows that the correlated contribution to the product of two luminosity marks scales as
\begin{equation}
\left\langle\delta m_i\delta m_j\right\rangle_r
\propto
\sigma_{\log L}^2 \rho^2
\xi_{ss}(r).
\label{eq:residual_mark_response}
\end{equation}

Thus, $\xi_{ss}(r)$ provides the scale-dependent term that distinguishes the two luminosity models: Model A contains only the mass-dependent marked signal and uncorrelated luminosity scatter, while Model B contains an additional correlated residual whose amplitude is controlled by $\sigma_{\log L}$ and $\rho$.

\section{Revealing property-based galaxy assembly bias}\label{sec:NullTest}

Models A and B define a controlled null experiment. Model A contains the luminosity dependence associated with halo mass together with independent scatter, whereas Model B preserves the same mean mass--luminosity relation and scatter while correlating the luminosity residuals with a secondary halo property.

This test describes a property-based galaxy assembly bias to model the dependence of galaxy properties (in this case luminosity) on secondary halo properties at fixed halo mass. 
This follows the broader definition in which galaxy assembly bias may arise when either the occupation or the properties of galaxies depend on halo properties beyond mass \citep{WechslerTinker2018}. In our construction, the galaxy occupation is held fixed in both models. Then for fixed number $N$, Model A luminosity $L$ satisfies
\begin{equation}
P_{\rm A}(L|M) = P(L|M),
\label{eq:model_A_probability}
\end{equation}
whereas for Model B
\begin{equation}
P_{\rm B}(L|M,s) \neq P(L|M).
\label{eq:model_B_probability}
\end{equation}

This differs from occupation-based implementations of galaxy assembly bias, in which a secondary halo property changes the probability distribution of galaxy occupation at fixed halo mass \citep{Zentner2014,Hearin2016,Contreras2021}. In our construction, the halo occupation and galaxy positions are held fixed. Model B changes the luminosity distribution conditioned on the secondary property, $P_{\rm B}(L\mid M,s,N)\neq P_{\rm A}(L\mid M,N)$, while preserving the mass-conditioned luminosity distribution after marginalizing over $s$, $P_{\rm B}(L\mid M,N)=P_{\rm A}(L\mid M,N)$. Thus, the unweighted two-point clustering is unchanged, and the difference between the models appears only in the luminosity-weighted statistic. Conceptually similar property-based constructions have been used in age-matching and galactic-conformity models, which correlate galaxy colour with halo age or concentration at fixed luminosity while leaving the underlying galaxy catalogue unchanged \citep{HearinWatsonDouglas2013,Paranjape2015}. Our construction applies this logic to luminosity residuals at fixed halo mass and isolates the resulting spatial signal using luminosity-weighted marked clustering.

Our primary observable is the difference between the projected marked correlation functions of the two models,
\begin{equation}
\Delta \mathcal{M}_p(r_p)
\equiv
\mathcal{M}_p^{\rm B}(r_p)
-
\mathcal{M}_p^{\rm A}(r_p).
\label{eq:DeltaMp_AB}
\end{equation}

For a fixed galaxy sample, Models A and B have identical galaxy positions. Their ordinary projected correlation functions are therefore identical by construction,
\begin{equation}
w_p^{\rm A}(r_p)
=
w_p^{\rm B}(r_p).
\label{eq:wp_A_equals_wp_B}
\end{equation}

A non-zero $\Delta \mathcal{M}_p(r_p)$ demonstrates that luminosity-weighted clustering contains information that is not present in the unweighted two-point correlation function.

A dependence of luminosity on a secondary halo property alone does not necessarily generate an additional clustering signal. The secondary property must itself be spatially correlated at fixed halo mass. In our model, these two ingredients are represented by the luminosity--secondary-property coupling $\rho$ and the marked correlation $\xi_{ss}(r)$, respectively. Their combination propagates the secondary halo dependence into the luminosity-weighted clustering signal. The expected leading-order response is then:
\begin{equation}
\Delta \mathcal{M}_p(r_p)
\simeq
\left[\sigma_{\log L}\rho\right]^2
\xi_{ss}(r_p),
\label{eq:AB}
\end{equation}

where $\xi_{ss}(r_p)$ is the marked correlation function of the standardized secondary halo property defined in Eq.~\eqref{eq:xi_ss}. The scale dependence of $\xi_{ss}$ determines the expected behaviour of the additional luminosity-weighted clustering signal.

If $s$ represents a halo property that can be standardised and it is related to halo assembly history, such as concentration, a detection of
\begin{equation}
\Delta \mathcal{M}_p(r_p) \neq 0
\label{eq:null_rejection_condition}
\end{equation}
implies that galaxy luminosity, and hence luminosity-weighted clustering, depends on halo information beyond mass alone. In the controlled experiment considered here, this additional dependence constitutes a galaxy assembly-bias term. 

Equation~\eqref{eq:AB} shows that the amplitude of the signal depends on luminosity scatter $\sigma_{\log L}$, the strength of the coupling $\rho$, and the clustering of the secondary property $\xi_{ss}(r_p)$. The marked auto-correlation is even in $\rho$ and primarily constrains the magnitude $|\rho|$ rather than its sign. Determining whether the luminosity residual correlates positively or negatively with the secondary halo property would require an observable that is linear in the two residual fields, such as their cross-correlation.

\section{Simulations and mock catalogues} \label{sec:Sims}
We use $N$-body simulations to characterize halo properties and test their possible contribution to luminosity-weighted clustering. We construct mock catalogues designed to resemble the DESI Bright Galaxy Survey (BGS) public data release 1 (DR1), a low-redshift spectroscopic galaxy survey whose targets are selected using photometry from the DESI Legacy Imaging Surveys \citep{Hahn2023,Dey2019}.

We construct DESI-BGS-like mock observations to test our luminosity-weighted clustering model. This analysis builds on earlier studies of luminosity- and morphology-dependent clustering using marked statistics in the Southern Sky Redshift Survey 2 \citep{Beisbart2000} and of luminosity- and colour-weighted marked correlations in the Sloan Digital Sky Survey Main Galaxy Sample \citep{Skibba2006,Sheth2005}. Our aim is to extend this line of work with a substantially larger low-redshift sample and to use the BGS-like mocks to test sensitivity to secondary halo properties. DESI-BGS is well suited to this purpose because it combines spectroscopic redshifts with broad-band photometry. We construct a volume-limited sample over the BGS redshift range following \citet{AbdulKarim2025}, applying $M_r<-21.5$ to the DR1 sample. This selection yields an approximately constant number density of $n\simeq5.5\times10^{-4}\,h^3\,{\rm Mpc}^{-3}$.

\subsection{Halo catalogue}
We create a mock sample from haloes of the catalogue from in \cite{Smith2017}, based in Millennium XXL simulations \citep{Angulo2012}, which provides a halo light-cone with enough number density to sample the target galaxies of DESI-BGS. Correspondingly, The Millennium XXL simulation runs in a box of size $L_{\rm box} = 3\,h^{-1}\ {\rm Gpc}$ with a particle resolution of $M_{\rm p}=6.17 \times 10^9\,h^{-1}\ M_{\odot}$. Due to requirements of the BGS targeted number density of DR1, we keep haloes of $M_{\rm h} = 10^{11}\,h^{-1}\ M_{\odot}$, which corresponds to roughly 16 particles. 

 We choose the halo catalogue instead of the galaxy mocks in \cite{Smith2024} to exploit extra properties beyond halo masses to define luminosity marks, which are not available in the released mock galaxy catalogues, and to have some flexibility with the galaxy prescription as presented in the next section. In particular, we use $r_{v_{\rm max}}$ to define concentration $c$ for individual haloes, which is present in the parent halo catalogue.

\subsection{Galaxy mock}

From the halo catalogue, we choose the HOD prescription presented in \cite{Smith2017}. Here, $N_{\rm cen}$ is a softened transition spline-like cumulative function using halo mass $M_{200m}$, so the parameters of central occupation like $\log M_{\min}$ and $\sigma_{\log M}$ are functions of a luminosity threshold that evolves with redshift to match the observed luminosity function. The mean satellite occupation is a power law above a threshold mass, with explicit cut-off where $N_{\rm sat}$ vanish below $M_0$. The central treatment is a bit different from the usual 5-parameter HOD definition in \cite{Zheng2005}, which is required to achieve the target number density of BGS light-cone with this halo catalogue. We refer the readers to the \cite{Smith2017,Smith2022} papers for a further explanation and how the luminosity function fit is performed. This HOD sets produce volume complete samples at a redshift range $0.1<z<0.4$, to mimic the same behaviour of BGS DR1. The initial galaxy number density using this method yields $n_{\rm gal} = 9.28 \times10^{-4}\,h^3\ {\rm Mpc}^{-3}$, which we randomly downsample to the DR1 target $n_{\rm gal} = 5.5\times10^{-4}\,h^3\ {\rm Mpc}^{-3}$. The main purpose is to map halos to galaxies with luminosities, include satellite galaxies, and take intra-halo correlation to a more realistic level than the CLF can provide. We use the average of the luminosity--mass relation provided by the \cite{Smith2022} BGS-mock, and use $\sigma_{\log L} = 0.23$ (or 0.15 DEX), which is the mean value of the catalogue. Although this value slightly varies for both for halo mass and redshift, we keep it fix as a first approach, then we vary it during the robustness test in Section~\ref{ch:robustness}.


\subsection{Abacus Summit suite}\label{ch:Abacus}

We include independent halo catalogues from the \texttt{small} suite of \texttt{Abacus Summit} simulations, which consist in 1803 simulations with $M_{\rm p} = 2.1\times10^{9}\,h^{-1}\ M_\odot$ in a box of $L_{\rm box} = 500\,h^{-1}\ {\rm Mpc}$, prepared to build covariance of the respective summary statistics ($w_p$ and $M_p$). These simulations have their own, but equivalent definitions of halo properties \citep{Hadzhiyska2022}. We do the calculations in the redshift snapshot at $z=0.35$ as a representative volume to measure sample variance in our statistics. We use the HOD parameters of \cite{Prada2025}, for a volume limited sample of $M_{r} < -21.5$. the HOD parameter set is $ \left\{\log M_{\min}, \log M_0, \log M_1, \sigma_{\log M}, \alpha \right\} = \left\{13.31, 11.95, 14.15, 0.73, 0.97\right\}$ reaching a number density $n_{\rm gal}= 4.8\times10^{-3},h^3,{\rm Mpc}^{-3}$ \cite{Prada2025,Fernandez-Garcia26}. These HOD is reported to be used with the \texttt{Uchuu} simulations \citep{Uchuu2021}, but we reach the same target number density as the volume-limited sample reported \cite{Fernandez-Garcia26}. To generate each mock, we use a new random seed for each simulation to quantify HOD random noise (see Section \ref{ch:robustness}), which represents a relevant stochastic term in the marker correlation function calculations. We also check the mean clustering and marked clustering over the covariance results match the ones from the production mock, reaching agreements below 3\%.

\section{marked correlation function in DESI BGS mock}\label{sec:Results}
\begin{figure}
    \centering
    \includegraphics[width=0.99\linewidth]{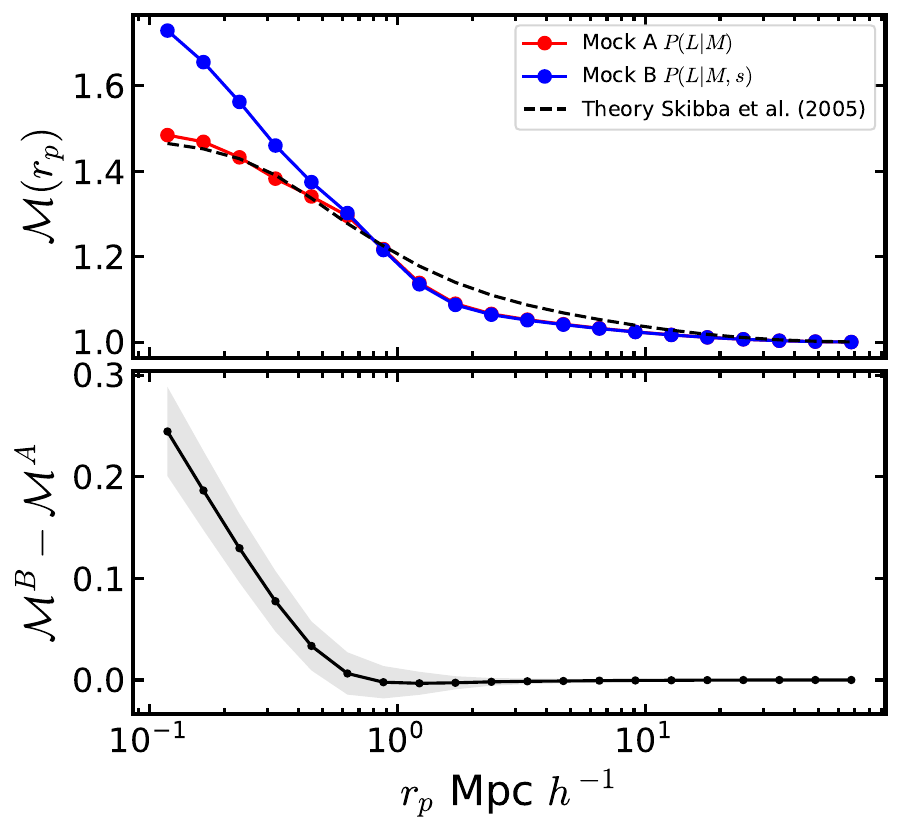}
    \caption{Top: Luminosity-weighted marked correlation function $\mathcal{M}_p$ as function of projected distance $r_p$ for BGS mocks of Model A (red) and Model B (blue). We add the theory described in Skibba et al. to show the the behaviour of the mass only model for lower amplitude of $\mathcal{M}_p$. Bottom: $\Delta \mathcal{M}_p\equiv \mathcal{M}_p^B - \mathcal{M}_p^A$ for the functions showed in top. We additionally add the standard deviation of the $\Delta \mathcal{M}_p$ data vector obtained from the covariance to show the variation around zero. The scale dependence of $\Delta \mathcal{M}_p$ is proportional to the marked correlation function of the secondary property $\xi_{ss}$.}
    \label{fig:DeltaMp_rp}
\end{figure}

\subsection{Estimator}

We estimate the correlation functions in redshift space using the \texttt{pycorr} package. The mock galaxies and random points positions are supplied as $(\mathrm{RA},\mathrm{Dec},\chi)$, where $\chi$ is the comoving distance in $h^{-1}\,\mathrm{Mpc}$ calculated using a flat $\Lambda$CDM cosmology with $H_0=73\,\mathrm{km\,s^{-1}\,Mpc^{-1}}$ and $\Omega_{\rm m}=0.25$. We use the \texttt{rppi} estimator, which bins pairs according to their transverse separation $r_p$ and line-of-sight separation $\pi$. The transverse-separation or projected distance bins are logarithmically spaced between $0.1$ and $80\,h^{-1}\,\mathrm{Mpc}$ using 20 bins, while the line-of-sight separation is divided into 80 linearly spaced bins between $-40<\pi / (\,h^{-1}\,\mathrm{Mpc})<40$ as specified in \texttt{pycorr}.

The random catalogues follow the same angular and radial selection as the mock galaxies. 
Random right ascensions are drawn uniformly in angle, while declinations are drawn uniformly in $\sin({\rm Dec})$ so that the points are uniformly distributed over solid angle. Random redshifts are drawn by sampling the redshift distribution of the selected mock galaxies. The random catalogues contain $N_{\rm r} =3$ times as many random points as galaxies, where $N_{\rm r}$ was kept between 3 and 10, showing a stable result. 

The unweighted correlation function is estimated with the Landy--Szalay
estimator,
\begin{equation}
\xi(r_p,r_\pi)
=
\frac{DD
      -2DR
      +RR}
     {RR},
\end{equation}
where $DD$, $DR$, and $RR$ are the normalized data--data,
data--random, and random--random pair counts.


For each galaxy, we define the luminosity mark as defined by Equation \ref{eq:luminosity_mark}. The mean luminosity value is $\langle L \rangle=1.829\times10^{10}\,h^{-2}\ L_\odot $, which is consistent with a magnitude range $-18.5\gtrsim M_r\gtrsim -23.8.$ observed by BGS galaxies with $M_r$ the rest frame $r$-band absolute magnitude. Luminosities are calculated from the luminosity--halo mass relation used in \cite{Skibba2006}, which is defined from the luminosity dependence in the correlation function from \cite{Zehavi2005}. This definition define the mark in the range $0.05 \lesssim m_i \lesssim 7.1$ with median $m_{\rm median}=0.831$. This means that the current definition of mark, up-weights more massive and luminous galaxies only, whereas around 60\% of galaxies have $m<1$. The marked correlation is measured by passing these marks as individual data weights to \texttt{pycorr}. The random catalogue is unweighted, corresponding to unit random weights, and the pair contribution to the weighted data--data and data--random counts is proportional to the product of the galaxy marks.
To asses the clustering of intra-halo galaxies where the luminosity effect is stronger, as schemed in Figure~\ref{fig:halo_luminosity_weighted_dependency}, we calculate the projected correlation function through
\begin{equation}
w_p(r_p)
=
\int_{-{\pi_{\max}}}^{+{\pi_{\max}}}
\xi(r_p,\pi)\,{\rm d}\pi.
\end{equation}

For the luminosity-weighted catalogue, the analogous quantity is calculated
from the weighted correlation function,
\begin{equation}
w_p^{\,L}(r_p)
=
\int_{-\pi_{\max}}^{+\pi_{\max}}
W_L(r_p,\pi)\,{\rm d}\pi.
\end{equation}
The projected marked correlation function is constructed from the ratio of the
projected luminosity-weighted and unweighted pair densities,
\begin{equation}
\mathcal{M}_p(r_p)
=
\frac{
\displaystyle
\int_{-\pi_{\max}}^{+\pi_{\max}}
\left[1+W(r_p,\pi)\right]\,{\rm d}\pi
}{
\displaystyle
\int_{-\pi_{\max}}^{+\pi_{\max}}
\left[1+\xi(r_p,\pi)\right]\,{\rm d}\pi
},
\end{equation}
The terms $1+\xi$ and $1+W$ represent the full unweighted and marked pair densities, respectively. The constant term accounts for the unclustered pair
contribution, while $\xi$ and $W$ describe the corresponding excess pair
counts. Alternatively, the projected marked correlation can equivalently be written as
\begin{equation}
\mathcal{M}_p(r_p)
=
\frac{
1+w_p^L(r_p)/(2\pi_{\max})
}{
1+w_p(r_p)/(2\pi_{\max})
},
\end{equation}
which accounts for the units ($[h^{-1}\, {\rm Mpc}]$) of the projected correlation function integrals.

\subsection{Measuring the secondary clustering term}
\begin{figure}
    \centering
    \includegraphics[width=0.94\linewidth]{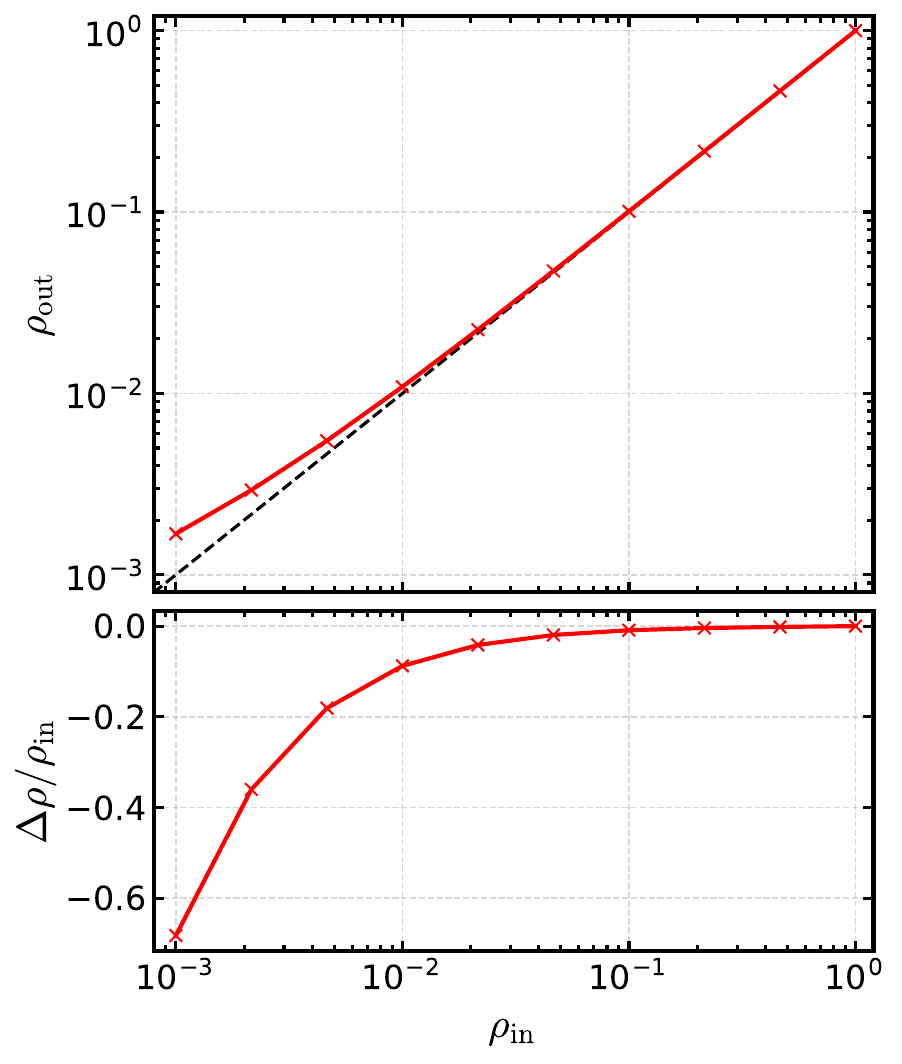}
    \caption{Top: Relation between the recovered $\rho_{\rm out}$ value vs the correlation coefficient used as the input $\rho_{\rm in}$ (red crosses). We show a range in logarithmic scale between $0.001 < \rho < 1$ to highlight the accuracy of the recovered value compared to the identity (black dashed line). Bottom: We add the relative residual $(\Delta\rho \equiv\rho_{\rm in} - \rho_{\rm out})/\rho_{\rm in}$ to highlight where the performance of the recovered $\rho_{\rm out}$ decreases, being less accurate ($<10\%$) for $\rho < 0.01$. This test indicates sub-percent accuracy for the explored range $0.1 < \rho < 1.0$.}
    \label{fig:rho_in_out}
\end{figure}

We estimate the luminosity-marked correlation function for the BGS mock we created and show it in Figure~\ref{fig:DeltaMp_rp}. The top panel of Figure~\ref{fig:DeltaMp_rp} shows the projected marked correlation function for mock generated cases A (red) and B (blue), where residuals are uncorrelated and correlated respectively. We also include a theory fit based on the Skibba et al. halo model, with best-fit parameters $\log_{10}(M_{\min}/h^{-1}M_\odot)=12.948$, $M_1/M_{\min}=48.89$, $\alpha_{\rm HOD}=1.218$, and satellite luminosity fraction $L_{\rm sat}/L_{\rm cen}=0.5$. The occupation fraction is $f_{\rm occ}=0.625$, giving $n_{\rm gal}=5.21\times10^{-4}\,h^3\,{\rm Mpc}^{-3}$. The theory yields a number density similar to the BGS mock, which validates the CLF model as a simpler approach for creating a mock where luminosity depends only on mass. For each halo of mass $M_i$, we use the individual concentration $c_i$ and then calculate $s_i$ using equation \eqref{eq:concentration_standardization}. We use $\log \bar{c}$ and $\sigma_{\log c}$ from a 40-bin configuration, where the first 30 bins are equally spaced in $\log M$ between $10^{11}$ -- $10^{14}\,h^{-1}M_\odot$. The remaining 10 bins cover the high-mass tail, $M>10^{14}\,h^{-1}M_\odot$, and are constructed to contain approximately the same number of haloes. 
In the bottom panel of Figure~\ref{fig:DeltaMp_rp}, we show $\Delta \mathcal{M}_p$ to test Equation~\eqref{eq:null_rejection_condition}, where we see a large absolute deviation consistently happening below $r< 1\,h^{-1}\ {\rm Mpc}$ reaching a maximum in the smallest tested distanced up to $r= 0.1\,h^{-1}\ {\rm Mpc}$. This result in an intra-halo term dependency that the marked correlation function difference $\Delta \mathcal{M}_p$ captures. We calculate uncertainties of $\Delta \mathcal{M}_p$, which are derived from the covariance matrix, using the \texttt{Abacus Summit small} set. The previous result represents the product $\sigma_{\log L}^2\rho^2\xi_{ss}$ as showed in Equation~\ref{eq:AB}. The scale dependence of $\Delta \mathcal{M}_p$ represents $\xi_{ss}$ the $s$-weighted-field correlation function. As this is a leading order result an independent calculation of $\xi_{ss}$, it can be used to recover the correlation coefficient amplitude $|\rho|$. We explain this in the next section.

\subsection{Test recovery}

To test whether this model can be applied to the real observable data, we look to recover the correlation coefficient $\rho$ by applying Eq.\eqref{eq:AB}. In this controlled experiment, $\sigma_{\log L}$ is known (from a calibrated simulation), and $s$ is infused in the luminosity mass relation correlated with coefficient $\rho_{\rm in}$. As we can estimate $\xi_{ss}$ independently through the mock, we fit $\rho_{\rm out}$ directly from our measurement of $\Delta \mathcal{M}_p$. Such a test intends to test the strength of the coupling between luminosity residual and $s$, and to reveal the minimum value of $\rho_{\rm out}$ we can recover. 

The potential detection of assembly bias by estimating $\Delta \mathcal{M}_p$ between theory (or mass-only dependent mock) and the observed luminosity marked correlation functions depends on the strength of the coupling, making the recovery of $\rho_{\rm out}$ a test to validate the model. For this purpose, we need two ingredients: A template to fit $\rho$ to the observed $\Delta \mathcal{M}_p$. Then, at fixed $\sigma_{\log L}=0.23$ as extracted from the BGS mock, we are trying to find the amplitude $\hat{q}$:

\begin{equation}
    \hat{q} = \frac{\xi_{ss}^{\intercal} C^{-1}\Delta \mathcal{M}_p}{\xi^{\intercal}C^{-1}\xi},
\end{equation}
that minimises the distance between $\xi_{ss}$ and $\Delta \mathcal{M}_p$. The least squares then finds $\rho_{\rm out} = \sqrt{q}/\exp{(\sigma_{\log L})}$. The recovered $\rho_{\rm out}$ is shown in Figure~\ref{fig:rho_in_out}, where for the tested values of $\rho_{\rm in}$, which is injected as shown in Eq.\eqref{eq:model_B}, are very close to the expected real value of $\rho$ in Figure~\ref{fig:rho_in_out}. This seems to be valid even for smaller values of $\rho$, where the precision of $\rho$ is lost at 10\% level for small values of $\rho\sim0.01$. However, we consider these $\rho$ values extremely low compared to the real cases we want to test, and these will also be consistent with a random case where errors are uncorrelated ($\rho=0.0$). We keep values of between $0.1 < \rho < 1.0$, which are all recovered with sub-percent precision to evaluate the possible detection of $\xi_{ss}$ in the data.

\begin{figure}
    \centering
    \includegraphics[width=0.95\linewidth]{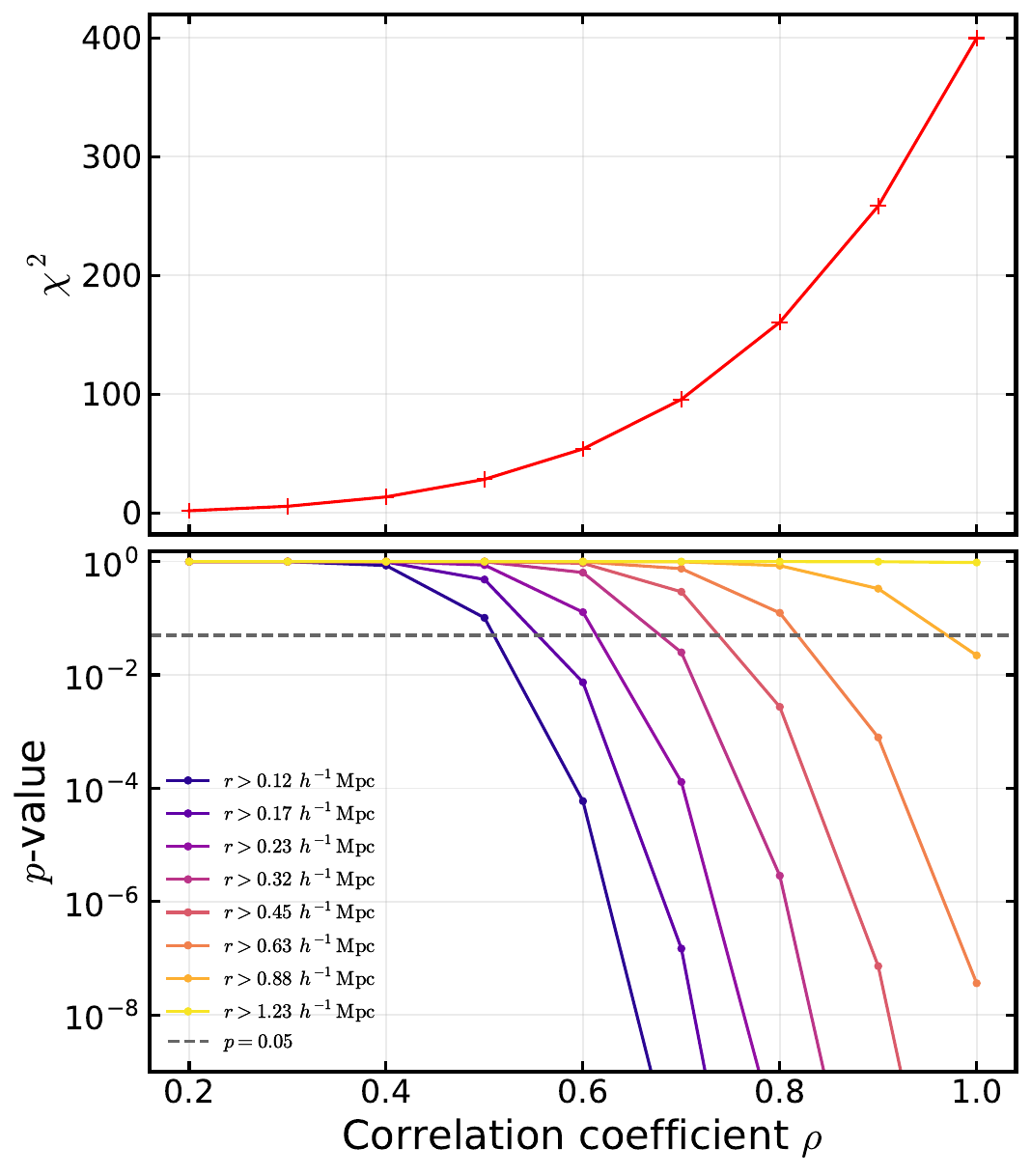}
    \caption{Top: $\chi^2$ value as function of the correlation coefficient $\rho$ for the range $0.2<\rho<1.0$ using 20 degrees of freedom. Bottom: the correspondent $p$--value for the $\chi^2$ values in top as function of $\rho$ and the number of distance bins in $\Delta \mathcal{M}_p$ displayed by the middle of the distance bin (colour lines). For example a detection ($5\sigma$ significance) occurs when $p\leq 2.13\times10^{-7})$, which is crossed for $\rho \approx0.64$. For higher values of $\rho$ detection is still calculated even for the removed smaller distance bins, where the information is contained. We also highlight the $2\sigma$ (indication; dashed line) where $\xi_{ss}$ could still be measured againts the null case.}
    \label{fig:chi2_p_value}
\end{figure}
\subsection{Detection, robustness and systematics} \label{ch:robustness}

To determine whether we can claim a detection from the null model, we need the following: Given the covariance of the null case (model A), what values of $\rho$ and up to what scale of $\xi_{ss}$ the $\chi^2$ defines a $p$-value where this effect indicates a measurement. The $\chi^2$ test is shown in Figure \ref{fig:chi2_p_value}, where we calculate this value as function of $\rho$ (top panel). As expected the $\chi^2$ value grows with the correlation strength being maximum at $\rho = 1$ (or $\rho = -1$). We emphasize that $|\rho|=1$ is far from realistic, because it would imply a full correlation between scatter of $L(M)$ and $c(M)$, which results in a deterministic relation between $L$ and $c$. For more realistic scenarios, we find that for $\rho>0.512$ the curve goes over valid p-value (dashed line, bottom panel) threshold using all the explored scales of $\Delta \mathcal{M}_p$ as shown in the bottom panel of Figure~\ref{fig:chi2_p_value}. At the same time we find that we only need scales of $r_p > 0.88\,h^{-1}\ {\rm Mpc}$ for large correlation or $\rho \sim 1$. Statistically, this means that for these configurations the test has less than 5\% to be null case. For a detection ($5\sigma$, using 20 degrees-of-freedom (dof)),  $\rho \geq 0.629 $ is required with all distance bins ($r>0.11\, h^{-1}\ {\rm Mpc}$) with $p=2.13\times 10^{-7}$.

 We perform additional test to quantify the impact of some of the choices made during the calculations of the marked correlation function to justify the detection of this galaxy-luminosity assembly bias through this method. We run 100 mock realizations with the fixed HOD parameters of the abacus run varying the random seeds that set halo occupation of central and satellite galaxies. For each realization, different haloes contribute to the mock, which introduces a possible selection effect in the mass-based mark as haloes will have different internal structure, as we look for differences in the 1-halo term. The projected correlation functions $w_p(r_p)$ and $\mathcal{M}_p$ are measured in the same default configuration than the production and covariance mocks. At scales where $r_p < 1h^{-1}$ Mpc, the scatter introduced by the HOD variance (we call it HOD noise) reaches a maximum of $5.5\%$ in the smallest distance bin. This is expected as we measure how the random noise of the HOD (variance) impacts the marked correlation function, which is an important factor, but it is still smaller than the marked correlation sample covariance, which is between 17-20\% as shown in Figure~\ref{fig:DeltaMp_rp}.

To determine how the luminosity--mass relation scatter changes as function of halo mass affecting the possible detection of galaxy assembly bias. We run the test for the fixed HOD varying $\sigma_{\log L}$. The first test includes a smaller value of $\sigma_{\log L}=0.1$ DEX, than the one used initially ($\sigma_{\log L}=0.15$ DEX). In the other hand, we include $\sigma_{\log L}=0.2$ DEX, to have an idea of the scaling of the detectability for different scatter amplitudes making the test for a fixed value of $\rho=0.7$. The test shows variations in the $\chi^2$ affecting the significance by $\sim2$--$3\sigma$, meaning that if we consider $\sigma_{\log L} = 0.1$ DEX with reduced luminosity scatter the detectability of the model is largely reduced. This would affect models with correlation strength of $\rho\sim0.5$. In particular, for the tested case with $\rho=0.7$, the significance is around $\sim4\sigma$ even after the reduced luminosity scatter is applied. For the case where $\sigma_{\log L}=0.2$ DEX, the detection improves, this is due to the observable $\Delta \mathcal{M}_p$ scaling directly with $\sigma_{\log L}^2$. In the real case the luminosity--mass scatter is both a function of halo mass and redshift, which fluctuates between $0.11 < \sigma_{\log L}<0.17$, making it necessary to accurately quantify this value for applying this test. We leave a varying modelling of $\sigma_{\log L}$ for a future study, where a real observational dataset is used.

\begin{figure}
    \centering
    \includegraphics[width=0.95\linewidth]{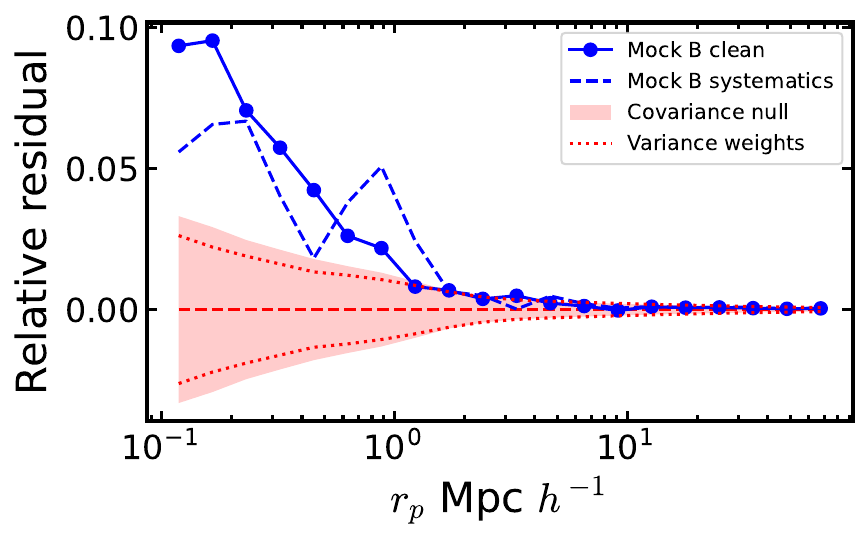}
    \caption{Relative residual $\Delta\mathcal{M}_p/\mathcal{M}_p^A$ as function of distance $r_p$ for the case with $\rho=0.7$ and $\sigma_{\log L} = 0.23$ ($0.15$ DEX) (blue) when systematic effects are accounted (dashed line) divided by the null case (model A; red dashed line). In this scenario, we account for galaxy selection systematics modelled with the Inverse Individual Probability (IIP) scheme, plus angular selection weight. We show the relative residual to highlight how $\Delta \mathcal{M}_p$ looks against the standard deviation from the null covariance (red shaded region) and how the variance of the weights is included (red dotted line).}
    \label{fig:systematic_marks}
\end{figure}

Additionally, we study the behaviour of $\Delta \mathcal{M}_p$ when systematic weights are included in the modelling. As the DESI clustering modelling reveals, PIP weights that account for the incompleteness of fibre collisions remain a relevant systematic to deal with, when studying the clustering at small scales \citep{Smith2019}, where more of the signal of this test comes from. By defining a luminosity-scale-dependent relation, it is relevant to correctly model the observation of galaxies inside the same halo. Considering, the complete addition of PIP weights requires resources we currently lack (mocks that actually account the spectrograph fibre patterns), we perform a simplified test using the individual inverse probability, which accounts for loosing galaxies at small separations. In summary, we introduce the redshift probability $p(z)$, which gives a weight to the individual galaxies depending on their redshifts. The current scheme removes between 20--35\% of galaxies in the explored redshift range with a small impact in both the $w_p$ and $\mathcal{M}_p$ calculation, producing between 1-4\% differences in the relevant scales. The results impacts $w_p$ more as the scheme is similar to a random sampling (removing $\sim30\%$), producing an additional bias of 1\%. We also account for angular selection, including a distance dependent weight which accounts for loosing galaxies due to fibre collisions. We decide to keep the current IIP+angular weights for the recovering, to account the additional introduced noise due to this particular systematic. We apply this for Mock A and Mock B and the case of $\sigma_{\log L} = 0.23$ and $\rho = 0.7$, to see how would affect a possible detection. In Figure~\ref{fig:systematic_marks}, we asses the impact of the new weights in our configuration, plotted as the relative residual with respect to the respective null case (with and without systematics). We see the effect of this weighting scheme in our test for this configuration impacts small scales and adds an extra variance on the same level than the null-covariance, reducing the final $\chi^2$--value by $1.6\sigma$. Such a effect would become relevant again for the cases where $\rho\sim0.5$, where the $p=0.05$ is close to the indication threshold. A more faithful approach to overcome this issue, would be to use the correct PIP case, which deals better with fibre collisions. We leave this analysis for a future work, which will be necessary to claim a detection. As this study points more to the proof-of-concept approach, where the intention is to study how the systematics affect the robustness of the test, we keep this simpler analysis.

\section{Conclusions} \label{sec:conclusions}

We use the luminosity-weighted marked correlation function to test whether galaxy luminosities contain spatially correlated secondary halo information beyond that encoded by halo mass. We describe a secondary halo property through its standardized residual \(s\) at fixed mass and characterize the new spatial correlation \(\xi_{ss}\). As a proof of concept, we correlate luminosity residuals with concentration residuals at fixed halo mass, using a concentration proxy derived from \(r_{v_{\max}}\). This construction preserves the mean luminosity--mass relation and its scatter while generating an additional, scale-dependent contribution to the marked correlation function. Although concentration provides a physically motivated example because of its connection to halo assembly history and secondary halo bias, the framework can be applied to any halo property that can be standardized at fixed mass and whose residuals are spatially correlated.

We apply the test to a mock catalogue that resembles the DESI-BGS DR1. Because \(\xi_{ss}\) can be measured independently in the simulation and \(\sigma_{\log L}\) is fixed, we recover the amplitude of the luminosity--secondary-property coupling, \(|\rho|\). In the baseline configuration, the additional marked-clustering contribution can be distinguished from the null hypothesis at 95\% confidence for \(|\rho|>0.52\), assuming \(\sigma_{\log L}=0.23\). Using a null covariance that includes sample variance and stochasticity in the halo occupation, the mass-only hypothesis is rejected at greater than \(5\sigma\) significance for \(|\rho|>0.63\) when scales \(r_p>0.11\,h^{-1}\,{\rm Mpc}\) are included. The sensitivity decreases when the smallest scales are removed, demonstrating that much of the constraining information arises from intra-halo galaxy pairs. Tests of HOD stochasticity, luminosity-scatter uncertainty, and approximate fibre-assignment and angular-selection effects show that these systematics can reduce the detection significance. A complete treatment using PIP weights will be required when applying the method to DESI observations.

The framework developed here can be applied directly to observational data by comparing the measured luminosity-weighted marked correlation function with a mass-only prediction calibrated to the same galaxy sample. A non-zero luminosity-marked signal alone would not constitute evidence for galaxy assembly bias, since such a signal is already expected from the joint dependence of galaxy luminosity and clustering on halo mass. The relevant diagnostic is instead a statistically significant, scale-dependent deviation from the calibrated mass-only baseline after its modelling uncertainties and observational systematics have been accounted for. Once uncertainties in this baseline and observational systematics are accounted, such a deviation would indicate that luminosity residuals contain spatially correlated information beyond halo mass, providing evidence for a property-based manifestation of galaxy assembly bias. The principal contribution of this work is therefore not a new marked statistic, but a controlled interpretation that turns an established measurement into an observational null test for property-based galaxy assembly bias.

\begin{acknowledgments}
JA acknowledges useful discussions with Tiago Castro, Daniel López-Cano, and Ginevra Favole. This work was supported by the São Paulo Research Foundation (FAPESP), Grant No. 2025/11898-5. RV was supported by a FAPESP fellowship, Proc. 2026/00876-3 . RA acknowledges support from FAPESP and CNPq. ADMD acknowledges support from the Universidad Técnica Federico Santa María through the Proyecto Interno Regular \texttt{PI\_LIR\_25\_04}. OpenAI ChatGPT and Codex Sol 5.6 were used to assist with language editing and proofreading. All suggestions were reviewed and revised by the authors.
\end{acknowledgments}

\bibliography{apssamp}

\end{document}